\documentclass [amssymb, amsmath, showpacs,nofootinbib]{revtex4}

\usepackage{graphicx}
\usepackage{dcolumn}
\usepackage{bm}
\usepackage[all]{xy}

\begin{document}

\title{Fermions and Loops on Graphs. I. Loop Calculus for Determinant}
\author {Vladimir Y. \surname{Chernyak}$^{a,b}$}
\author{Michael \surname{Chertkov}$^{b}$}

\affiliation{$^a$Department of Chemistry, Wayne State University,
5101 Cass Ave,Detroit, MI 48202}

\affiliation{$^b$Center for Nonlinear Studies and Theoretical Division, LANL, Los Alamos, NM 87545}

\date{\today}

\begin{abstract}
This paper is the first in the series devoted to evaluation of the partition function in
statistical models on graphs with loops in terms of the Berezin/fermion integrals. The paper
focuses on a representation of the determinant of a square matrix
in terms of a finite series, where each term corresponds to a loop on the graph. The representation
is based on a fermion version of the Loop Calculus, previously introduced by the authors for
graphical models with finite alphabets. Our construction contains two levels. First, we represent
the determinant in terms of an integral over anti-commuting Grassman variables, with some
reparametrization/gauge freedom hidden in the formulation. Second, we show that a special choice of
the gauge, called BP (Bethe-Peierls or Belief Propagation) gauge, yields the desired loop
representation. The set of gauge-fixing BP conditions is equivalent to the Gaussian BP equations,
discussed in the past as efficient (linear scaling) heuristics for estimating the covariance of a
sparse positive matrix.
\end{abstract}



\maketitle

The series in general, and this  paper in particular, belongs to the new emerging field of
statistical inference and graphical models born on the cross-roads of statistical/mathematical
physics, computer science, and information theory (see the following recent books as introductory
reviews \cite{02HR,03Mac,08MM}). A typical problem in the field can be stated as follows. Given a
graph (trees, sparse graphs or lattices are the three most popular examples), finite or infinite
alphabet with the variables defined on the graph elements (typically vertexes or edges) and a cost
function (probability) associated with any given variables configuration, one should find the
marginals, correlation functions, or solve the weighted counting problem (calculate the partition
function). The main problems that define the field are: (a) to estimate the efficiency of exact
evaluation for a typical or worst case problem for a class in terms of its dependence on the
problem size; and (b) when an exact evaluation is not feasible, as it requires an unacceptably
large number of steps, to suggest an approximation and a corresponding efficient algorithmic
implementation.

A powerful approach in the field of statistical inference is to build an efficient scheme based on
a simple case or limit, where the evaluation is easy, i.e. complexity is polynomial in the number
of variables. One simple case corresponds to trees, i.e. graphs without loops. Following the
so-called Bethe-Peierls approach inspired by \cite{35Bet,36Pei}, one can show that the
computational effort for the partition function on a tree is linear in its size. Furthermore, one
anticipates that the tree-based methods and corresponding algorithms should perform reasonably well
on sparse graphs with relatively few loops.  This approach was reinvented and successfully explored
in coding theory \cite{63Gal} (see e.g. \cite{08RU} for a modern discussion of the graph-based
codes) and artificial intelligence \cite{88Pea},  where the corresponding algorithm was coined
Belief Propagation (BP) and this name is now commonly accepted across the disciplines. In a recent
development we suggested an approach, called Loop Calculus (LC) \cite{06CCa,06CCb}, which
establishes an explicit relation between the BP (previously thought of as just heuristics) and
exact results. Formally, LC expresses the partition function of a graphical model in terms of a
series over certain subgraphs (referred to as generalized loops), where each individual term (that
corresponds to a generalized loop) is expressed explicitly in terms of the BP solution (strictly
speaking, a fixed point of the corresponding BP equations). LC, originally formulated for a binary
alphabet, has been extended to an arbitrary finite alphabet in \cite{07CC}, and the corresponding
approach has been called the Loop Tower.

There is also a class of problems that are easy in spite of a large number of loops contained in
the underlying graphical structure. A so-called Gaussian Graphical Model (GGM) that belongs to this
class is closely related to the subject of this paper. Consider a graphical model with continuous
variables defined on vertexes of a graph with a Gaussian pair-wise interaction. The corresponding
partition function, represented simply by a multi-dimensional Gaussian integral, is, therefore,
reduced to evaluating the covariance (inverse) of the interaction matrix (note that this operation
is well defined only if the matrix is positive definite). For an arbitrary interaction matrix,
expressed in terms of a dense graph, this is a problem of $N^3$ complexity, where $N$ is the number
of the graph vertices. However, as shown in \cite{01WF,01RV}, one can also use a more efficient,
linear in $N$, Gaussian Belief Propagation (GBP) algorithm for exact calculations of the marginals
within the Gaussian continuous-alphabet model. The GBP can also be used for finding the covariance
or evaluating the determinant of the interaction matrix. However, it does not give an exact result,
but rather provide an approximation, which is conjectured to be a reasonably accurate heuristics at
least for sufficiently sparse graphs. An intuitive (but also mathematically rigorous) explanation
for the exactness of GBP in the case of marginals and its insufficiency for the covariance and
related object has been given in \cite{06MJW} via the so-called Walk-Sum Approach (WSA). WSA
relates the exact result for the covariance to the sum over all possible oriented paths on the
graph, while GBP (strictly speaking asymptotic GBP, evaluated on an infinite computational tree)
corresponds to the summation over a special sub-family of directed walks, specifically backtracking
directed walks. This approach has also been extended to evaluate the partition function of GGM
(related to the determinant of the interaction matrix) in \cite{08Joh}. The majority of these and
other recent studies of GGM have focused on the analysis of the conditions for the Gaussian BP
convergence \cite{01RV,03WJW,05RH,06MJW,08MV,08CH} or practical implementations of the GBP
algorithm \cite{05RH,08SSWBD}.

However, and in spite of this important progress and practical significance, a systematic analysis
of the accurateness of GBP as an approximation and a possibility for systematic improvements of the
GBP so far has been largely left unexplored. Even though we are still unable to provide the full
answers, this paper reports some progress towards the future resolution of these important
questions.

In this paper we introduce a fermion-based version of the Loop Calculus approach of
\cite{06CCa,06CCb,07CC} that provides an exact representation for a determinant as a finite loop
series, where the first term corresponds to a fixed point of the GBP algorithm for the
corresponding GGM (that we will also be calling a solution of the GBP equations, or simply GBP
solution). Each subsequent term of the Loop Series is associated with a loop on the underlying
graph and is expressed explicitly in terms of the GBP solution. Our approach explores the power of
the Berezin representation for a determinant in terms of symbolic integrals over anti-commuting
Grassman variables \cite{87Ber,80FS}. Note, that a relation between some binary graphical models on a
planar graph and Grassman integrals was briefly discussed in \cite{08CCT}. In this and subsequent
papers of the series \cite{08CCb} we take broader perspectives and do not limit our discussions to
planar graphs.

The paper is organized as follows. In Section \ref{sec:det-integral} we start our discussion with
an extensive introduction (reminder) to the Berezin integral approach, that will culminate in a
Grassman integral representation for the determinant of the underlying correlation matrix. Section
\ref{sec:Loop-series-det} is split in three Subsections and forms the core of the paper. Gauge
transformations that keep the partition function of the fermion model (the determinant) invariant
are introduced in Section \ref{subsec:Gauge}. In Section \ref{subsec:BP-equations} the Belief
Propagation is interpreted as a gauge fixing condition. Section \ref{subsec:loop-series} finalizes
the construction of the Loop Series. The manuscript also contains two Appendices. Appendix
\ref{sec:node-edge-factors} is auxiliary to Section \ref{subsec:loop-series} and contains some
technical details of the Grassman integral calculations. Appendix \ref{sec:bose} derives the BP
equations for the standard (Bose) representation. Section \ref{sec:summary} summarizes the results,
briefly discusses the relations between the results reported in this paper and other results and
future directions, e.g. related to the second paper in the series \cite{08CCb}.

\section{Introduction: Integral Representation}
\label{sec:det-integral}

We start with introducing a convenient integral representation for the determinant of an $N\times
N$ matrix $H_{ab}$ that will allow to apply the Loop Calculus approach \cite{06CCa,06CCb} to
represent $\det H$ in terms of a finite loop series. Although formally the proposed scheme can be
applied to any matrix, it becomes algorithmically  practical in the case when the matrix $H$ is
sparse.

We associate with our $N\times N$ matrix $H$ a graph ${\cal G}(H)$ with $N$ nodes $a\in{\cal
G}_{0}$. The nodes $a$ and $b$ are connected by edge $\alpha=\{a,b\}\in{\cal G}_{1}$, where $a\neq b$, when
$H_{ab}\ne 0$, or $H_{ba}\ne 0$. In other words the nodes $a\in{\cal G}_{0}$ represent the diagonal
elements $H_{aa}$, whereas the edges $\{a,b\}\in{\cal G}_{1}$, correspond to non-zero off diagonal
matrix elements $H_{ab}$ and $H_{ba}$. Hereafter we will also use a convenient notation $a\sim b$
for $\{a,b\}\in{\cal G}_{1}$. To avoid confusion we emphasize that $\{a,b\}$ naturally denotes a
set that consists of (two) elements $a$ and $b$, i.e., $\{a,b\}\in 2^{{\cal G}_{0}}$ is a subset of
the vertex set ${\cal G}_{0}$. In particular $\{a,b\}=\{b,a\}$, which means that we are dealing
with a non-oriented graph, and ${\cal G}_{1}\subset {\cal G}_{0}^2$ denotes the set of graph
(non-oriented) edges. The notation $(a,b)$ with $a\neq b$ stands for ordered pairs, i.e.
$(a,b)\ne(b,a)$. Ordered pairs can be utilized to denote oriented edges $(a,b)\in{\cal
G}_{0}\times{\cal G}_{0}$, if we decide to choose an orientation on our non-oriented graph. Note
that the sparseness of $H$ means that the valences (degrees of connectivity at nodes), ${\rm
val}(a)={\rm card}(\{b\in{\cal G}_{0}(H)|b\sim a\})$, are small compared to $N$.

To develop a finite loop decomposition for the determinant we represent $\det H$ as a
Berezin integral over anti-commuting Grassman variables \cite{87Ber}. Specifically, we
introduce a set $\{\bar{\bm\theta},{\bm\theta}\}=\{\bar{\theta}_{a},\theta_{a}\}_{a\in{\cal
G}_{0}}$ with $a=1,\ldots,N$ of Grassman variables that anti-commute, i.e.,
\begin{eqnarray}
\label{Grassman-variables} \theta_{a}\theta_{b}=-\theta_{b}\theta_{a}, \;\;\;
\bar{\theta_{a}}\theta_{b}=-\theta_{b}\bar{\theta}_{a}, \;\;\;
\bar{\theta_{a}}\bar{\theta_{b}}=-\bar{\theta_{b}}\bar{\theta_{a}}, \;\;\; \forall a,b \in{\cal
G}_{0}.
\end{eqnarray}
A function $F(\bar{\bm\theta},{\bm\theta})$ of the Grassman variables is understood as a Taylor
series, which turns out to be finite since, due to the anti-commuting relations
Eq.~(\ref{Grassman-variables}), each term of the Taylor series can contain any component
$\theta_{a}$ or $\bar{\theta}_{a}$ not more than once. The Berezin integral is defined via the
Berezin measure
\begin{eqnarray}
\label{Berezin-measure} {\cal D}{\bm\theta}{\cal D}\bar{\bm\theta}=\prod_{a\in{\cal
G}_{0}}d\theta_{a}d\bar{\theta}_{a},
\end{eqnarray}
where the differential variables $\{d\theta_{a},d\bar{\theta}_{a}\}_{a\in{\cal G}_{0}}$
anti-commute with each other and with the original Grassman variables. The Berezin measure is fully
defined by the integration rules
\begin{eqnarray}
\label{Berezin-int-rules} \int\theta_{a}d\theta_{a}=\int\bar{\theta}_{a}d\bar{\theta}_{a}=1, \;\;\;
\int d\theta_{a}=\int d\bar{\theta}_{a}=0.
\end{eqnarray}
For those who seek more rigorous definitions: We introduce a Grassman algebra ${\rm Gr}({\cal
G}_{0})$ as an algebra over $\mathbb{R}$ (or $\mathbb{C}$) generated by
$\{\bar{\theta}_{a},\theta_{a}\}_{a\in{\cal G}_{0}}$ with the relations given by
Eq.~(\ref{Grassman-variables}). A function $F$ of Grassman variables should be interpreted as an
element $F\in {\rm Gr}({\cal G}_{0})$ of the Grassman algebra. The Berezin integral is a measure
$\mu:{\rm Gr}({\cal G}_{0})\to \mathbb{R}$ that associates with any element of the Grassman
algebra(or simply a ``function of Grassman variables'') the value of its integral, according to the
rules given by Eqs.~(\ref{Berezin-measure}) and (\ref{Berezin-int-rules}).

A well-known property of the Gaussian Berezin integrals, applied to our case, reads
\begin{eqnarray}
\label{det-int-represent} \det H=\int{\cal D}{\bm\theta}{\cal
D}\bar{\bm\theta}e^{S_{0}(\bar{\bm\theta},{\bm\theta})}, \;\;\;
S_{0}(\bar{\bm\theta},{\bm\theta})=\sum_{a\in{\cal
G}_{0}}H_{aa}\bar{\theta}_{a}\theta_{a}+\sum_{a,b\in{\cal G}_{0}}^{a\sim
b}H_{ab}\bar{\theta}_{a}\theta_{b}.
\end{eqnarray}
According to Eq.~(\ref{det-int-represent}) we interpret the determinant as partition function of a
statistical fermion model defined on the graph ${\cal G}$, where the fermion (Grassman) variables
reside on the graph nodes. To apply the loop decomposition we convert the resulting statistical
model into a vertex model, i.e., the one with the variables residing on the graph edges. This task
can be easily accomplished with the help of a Hubbard-Stratanovich (HS) type transformation,
defined as follows. We introduce a set of Grassman variables
$\{\bar{\chi}_{ab},\chi_{ab}\}_{a,b\in{\cal G}_{0};\{a,b\}\in{\cal G}_{1}}$, describing the HS
decoupling field representing the off-diagonal terms in the action of Eq.~(\ref{det-int-represent}).
These variables express interaction of the original variables with the decoupling field.
This is achieved by making use of the set of identities
\begin{eqnarray}
\label{HS-identities}
e^{H_{ab}\bar{\theta}_{a}\theta_{b}+H_{ba}\bar{\theta}_{b}\theta_{a}}&=&-H_{ab}H_{ba}\int
d\chi_{ab}d\bar{\chi}_{ab}d\chi_{ba}d\bar{\chi}_{ba}e^{(H_{ab})^{-1}\bar{\chi}_{ab}\chi_{ba}
+(H_{ba})^{-1}\bar{\chi}_{ba}\chi_{ab}} \nonumber \\ &\times&
e^{\chi_{ba}\bar{\theta}_{a}+\bar{\chi}_{ba}\theta_{a}
+\chi_{ab}\bar{\theta}_{b}+\bar{\chi}_{ab}\theta_{b}}, \;\;\; \forall\{a,b\}\in{\cal G}_{1},
\end{eqnarray}
where, for the sake of simplicity, we assume that, if $a\sim b$, both matrix elements $H_{ab}$ and
$H_{ba}$ are non-zero. (The latter condition can be actually relaxed, however this goes beyond the
scope of this manuscript.) Using the integral representation (\ref{HS-identities}) for the terms
that originate from the off-diagonal terms of the action $S_{0}$ in Eq.~(\ref{det-int-represent})
we arrive at the following HS representation for the determinant
\begin{eqnarray}
\label{det-int-represent-2} \det H&=&\left(\prod_{\{a,b\}\in{\cal G}_1}(-H_{ab}H_{ba})\right)\int{\cal
D}{\bm\theta}{\cal D}\bar{\bm\theta}{\cal D}{\bm\chi}{\cal
D}\bar{\bm\chi}e^{S_{HS}(\bar{\bm\theta},{\bm\theta};\bar{\bm\chi},{\bm\chi})}, \nonumber \\
S_{HS}(\bar{\bm\theta},{\bm\theta};\bar{\bm\chi},{\bm\chi})&=&\sum_{a\in{\cal
G}_{0}}H_{aa}\bar{\theta}_{a}\theta_{a}+\sum_{a,b\in{\cal G}_{0}}^{a\sim
b}(H_{ab})^{-1}\bar{\chi}_{ab}\chi_{ba}+\sum_{a\in{\cal G}_{0}}\sum_{b\in{\cal G}_{0}}^{b\sim
a}(\chi_{ba}\bar{\theta}_{a}+\bar{\chi}_{ba}\theta_{a}).
\end{eqnarray}
As it always happens for the HS transformation, integration over the HS field in
Eq.~(\ref{det-int-represent-2}) reproduces the original integral representation
(\ref{det-int-represent}) due to the identities (\ref{HS-identities}). To accomplish the HS trick
we integrate over the original variables $\{\bar{\bm\theta},{\bm\theta}\}$. This can be readily
done, since the integration is local (i.e., can be performed on each node independently). Finally,
we arrive at the following desired expression for the determinant in a form of the partition
function of a vertex model
\begin{eqnarray}
\label{det=Z-vertex} \det H=Z=\left(\prod_{\{a,b\}\in{\cal G}_1}(-H_{ab}H_{ba})\right)
\left(\prod_{a\in{\cal G}_{0}}H_{aa}\right)\int{\cal D}{\bm\chi}{\cal
D}\bar{{\bm\chi}}\prod_{a\in{\cal
G}_{0}}f_{a}(\bar{{\bm\chi}}_{a},{\bm\chi}_{a})\prod_{\alpha\in{\cal
G}_{1}}g_{\alpha}(\bar{{\bm\chi}}_{\alpha},{\bm\chi}_{\alpha})
\end{eqnarray}
where ${\bm\chi}_{a}=\{\chi_{ba}\}_{b\in{\cal G}_{0},b\sim a}$ is a set of edge variables attached
to the node $a$, and ${\bm\chi}_{\alpha}=\{\chi_{ab},{\chi_{ba}}\}$ is the set of variables that
reside on edge $\alpha=\{a,b\}$. The edge functions
\begin{eqnarray}
\label{g-scalar-prod} g_{\alpha}(\bar{{\bm\chi}}_{\alpha},{\bm\chi}_{\alpha})
=g_{ab}(\bar{{\bm\chi}}_{\alpha},{\bm\chi}_{\alpha})=g_{ba}(\bar{{\bm\chi}}_{\alpha},{\bm\chi}_{\alpha})
=e^{(H_{ab})^{-1}\bar{\chi}_{ab}\chi_{ba} +(H_{ba})^{-1}\bar{\chi}_{ba}\chi_{ab}}, \;\;\;
\alpha=\{a,b\}
\end{eqnarray}
define the proper scalar products of the local states that belong to the same edge and different
nodes. The vertex factor-functions are obtained via the local integrations described above
\begin{eqnarray}
\label{f-vertex-functions} f_{a}(\bar{{\bm\chi}}_{a},{\bm\chi}_{a})=(H_{aa})^{-1}\int
d\theta_{a}d\bar{\theta}_{a}e^{H_{aa}\bar{\theta}_{a}\theta_{a}+\sum_{b\in{\cal G}_{0}}^{b\sim
a}(\chi_{ba}\bar{\theta}_{a}+\bar{\chi}_{ba}\theta_{a})}=e^{(H_{aa})^{-1}\sum_{b\in{\cal
G}_{0}}^{b\sim a}\bar{\chi}_{ba}\sum_{b'\in{\cal G}_{0}}^{b'\sim a}\chi_{b'a}}
\end{eqnarray}

\section{Gauge Transformation, Belief Propagation  and Loop Series}
\label{sec:Loop-series-det}

This Section is broken in three Subsections. In Section \ref{subsec:Gauge} we introduce a freedom
(gauge) allowing to represent an edge $g$-function as a sum of four terms each constituting a
product of two vertex terms. Section \ref{subsec:BP-equations} introduces a way of fixing the gauge
freedom in accordance with the Belief Propagation Principle. We further derive the Fermi-BP
equations which turn out to be fully equivalent to the (standard) Bose-BP equations discussed in
Appendix \ref{sec:bose}. The last Subsection \ref{subsec:loop-series} culminates in a derivation of
a finite Loop Series representation for the determinant, with each term of the series expressed
explicitly via the solution of the BP equations.

\subsection{Gauge Transformation}
\label{subsec:Gauge}

The $g$-terms in the integrand of Eqs.~(\ref{det=Z-vertex}) mix contributions associated with
different vertices. Our next step will aim at the decomposing of the $g$-contribution at any edge
into a sum of bi-local expressions. Specifically, we will be seeking for a decomposition of the
following type:
\begin{eqnarray}
\label{skew-orthogonality}
g_{ab}=c_{ab}\left(e^{\gamma_{ab}\bar{\chi}_{ab}\chi_{ab}}e^{\gamma_{ba}\bar{\chi}_{ba}\chi_{ba}}
+\kappa_{ab}e^{\gamma'_{ab}\bar{\chi}_{ab}\chi_{ab}}e^{\gamma'_{ba}\bar{\chi}_{ba}\chi_{ba}}\right)
+\zeta_{ab}\bar{\chi}_{ab}\chi_{ba}+\zeta_{ba}\bar{\chi}_{ba}\chi_{ab},
\end{eqnarray}
with $\kappa_{ab}=\kappa_{ba}$ and $c_{ab}=c_{ba}$. All the newly introduced parameters in
Eq.~(\ref{skew-orthogonality}) are to be defined by comparison with Eq.~(\ref{g-scalar-prod}).
Therefore, expanding Eq.~(\ref{skew-orthogonality}) and Eq.~(\ref{g-scalar-prod}) into a series
over the Grassman variables and comparing the results term by term one establishes the following
relations for the coefficients entering Eq.~(\ref{skew-orthogonality})
\begin{eqnarray}
\label{coeff-relations} c_{ab}(1+\kappa_{ab})=1, \;\;\; \gamma_{ab}+\kappa_{ab}\gamma'_{ab}=0,
\;\;\; \gamma_{ba}+\kappa_{ab}\gamma'_{ba}&=&0, \;\;\; \zeta_{ab}=(H_{ab})^{-1}, \;\;\;
\zeta_{ba}=(H_{ba})^{-1}, \nonumber \\
c_{ab}(\gamma_{ab}\gamma_{ba}+\kappa_{ab}\gamma'_{ab}\gamma'_{ba})&=&-(H_{ab}H_{ba})^{-1}
\end{eqnarray}
The relations (\ref{coeff-relations}) allow all the coefficients to be expressed in terms
of $\gamma_{ab}$:
\begin{eqnarray}
\label{coeff-expressions} \kappa_{ab}=-H_{ab}H_{ba}\gamma_{ab}\gamma_{ba}, \;\;\;
c_{ab}=\frac{1}{1+\kappa_{ab}} \;\;\;
\gamma'_{ab}=-(\kappa_{ab})^{-1}\gamma_{ab}=\frac{1}{H_{ab}H_{ba}\gamma_{ba}}, \;\;\;
\zeta_{ab}=(H_{ab})^{-1}.
\end{eqnarray}
We call the freedom in choosing the $\gamma$ variables the gauge freedom, as any choice of $\gamma$
does not change the value of the partition function defined by
Eqs.~(\ref{det=Z-vertex},\ref{g-scalar-prod},\ref{f-vertex-functions},\ref{skew-orthogonality},\ref{coeff-expressions}).

The gauge transformation, formally described above, utilizes a decomposition of the graphical trace
over the vector spaces $V_{ab}$, where each $V_{ab}$ is associated with the link $\{a,b\}\in{\cal
G}_{1}$ and node $b\in{\cal G}_{0}$: $V_{ab}$ represents the Grassman algebra generated by
$\bar{\chi}_{ab}$ and $\chi_{ab}$. It has a super-dimension $(2|2)$, i.e., an even dimension $2$
(the first number) and odd dimension $2$ (the second number), since it has $2$ even states, namely
$1$ and $\bar{\chi}_{ab}\chi_{ab}$, and $2$ odd states, namely $\bar{\chi}_{ab}$ and $\chi_{ab}$.

The skew-orthogonality conditions (\ref{skew-orthogonality},\ref{coeff-expressions}) are
different from these we have introduced in the context of the finite-alphabet graphical models
\cite{06CCa,06CCb,07CC}. Indeed, since the local state spaces $V_{ab}$ with the super-dimension
$(2|2)$ have total dimension $4=2+2$, a possible approach would be to build a tower hierarchy in
the spirit of \cite{07CC}. Then, a more general set of the skew-orthogonality conditions (compared
to those discussed above) should be introduced. However, the additional symmetry, i.e. the
superstructure (strictly speaking it should be referred to as a $\mathbb{Z}_{2}$-graded structure)
of the underlying linear algebra problem defines our choice of the more stringent
skew-orthogonality constraint. The details of (and actual reasons for) the choice will become
clear in the next Subsection when we discuss the additional BP constraints for the gauges.

The gauge transformation results in an explicit representation for the whole partition function in
terms of a series where each term corresponds to a choice of one of the four aforementioned states
at each edge. The expansion is derived via a direct substitution of
Eqs.~(\ref{skew-orthogonality},\ref{coeff-expressions}) into Eq.~(\ref{g-scalar-prod}), followed by
expanding the expression into monoms, substituting it back into the integrand of
Eq.~(\ref{det=Z-vertex}) followed by the evaluation of the resulting integrals term by term. Each
of the elementary integrals is vertex-local, thus turning the expression under evaluation into a
product of simple vertex-related contributions. For any choice of the edge local parameters
$\gamma$ one expects a gauge dependence of the individual contributions to the resulting series for
the partition function, while the cumulative result (the entire sum) will be gauge
insensitive/invariant by construction.

In the following Subsection we discuss a special choice of the gauge, related to the BP approach,
which essentially restricts all the contributions in the aforementioned series over the edge-states
to those that correspond to generalized loops (to be defined later) on the graph. Note that another
(non-BP) choice of the gauge that leads to an interesting explicit expression for the partition
function of the monomer-dimer model as a finite series expansion over determinants (which is also a
loop series of a kind, but in another sense) is discussed in the second paper of the series
\cite{08CCb}.

\subsection{Belief-Propagation Equations}
\label{subsec:BP-equations}

In this Subsection we extend our general approach, coined Loop Calculus \cite{06CCa,06CCb}, to the
Grassman integral for the partition function (determinant) of the Gaussian model.

We will impose additional constraints on the $\gamma$-gauges following two complementary
approaches. First, we describe the BP gauge as a result of the ground state optimization. Then, we
derive the same BP equations as a set of the no-loose-end constraints on the excited states.

\subsubsection{Variational derivation of BP equations}

The ground state contribution to the partition function (determinant) is naturally given by
\begin{eqnarray}
\label{BP-contribution} Z_0=\left(\prod_{\{a,b\}\in{\cal G}_1}\frac{H_{ab}H_{ba}}{H_{ab}H_{ba}
\gamma_{ab}\gamma_{ba}-1}\right)\left(\prod_{c\in{\cal G}_{0}}H_{cc}z_{c}\right), \;\;\; z_{c}=\int
d{\bm\chi}_{c}d\bar{{\bm\chi}}_{c}f_{c}(\bar{{\bm\chi}}_{c},{\bm\chi}_{c})\prod_{a'\sim
c}e^{\gamma_{a'c}\bar{\chi}_{a'c}\chi_{a'c}},
\end{eqnarray}
where the dependence on the $\gamma$-gauges is spelled out explicitly. Since
Eq.~(\ref{BP-contribution}) interprets $z_c$ as a Gaussian integral over the Grassman variables
associated with vertex $c$, a direct evaluation of the integral results in $z_c=\det(M_c)$, where
each element of the newly introduced matrix $M_c$ is defined by
$M_{c,ab}=(H_{cc})^{-1}+\gamma_{bc}\delta_{ab}$. Evaluating the determinant explicitly one finds
\begin{eqnarray}
\label{det-M} z_c=\det(M_c)=\left(\prod_{a'\sim
c}\gamma_{a'c}\right)\left(1+(H_{cc})^{-1}\sum_{a'\sim c}(\gamma_{a'c})^{-1}\right),
\end{eqnarray}
and the resulting expression for the ground state contribution adopts a form
\begin{eqnarray}
\label{Z0} Z_0=\left(\prod_{\{a,b\}\in{\cal G}_1}\frac{H_{ab}H_{ba}}{H_{ab}H_{ba}
\gamma_{ab}\gamma_{ba}-1}\right)\left(\prod_{c\in{\cal G}_{0}}H_{cc}\left(1+(H_{cc})^{-1}\sum_{a'\sim
c}(\gamma_{a'c})^{-1}\right) \left(\prod_{a''\sim c}\gamma_{a''c}\right)\right).
\end{eqnarray}

Considering $Z_0(\gamma)$ as a $\gamma$-dependent approximation for the full partition function
(by construction the latter does not depend on $\gamma$) one can define the BP-conditions
as an adjustment of $\gamma$ that minimizes the dependence of $Z_0$ on it. Formally, one looks for
a stationary point of $Z_0(\gamma)$:
\begin{eqnarray}
\forall a\in{\cal G}_0\ \ \mbox{and}\ \ \{a,b\}\in{\cal G}_1:\quad \left.\frac{\partial
Z_0}{\partial \gamma_{ab}}\right|_{\gamma^{\mbox{\small bp}}}=0\quad\Rightarrow \quad
-H_{ab}H_{ba}\gamma_{ba}^{(\mbox{\small bp})}=H_{bb}+\sum_{a'\sim b}^{a'\ne a}(\gamma_{a'b}^{(\mbox{\small bp})})^{-1}.
\label{BP_eqs}
\end{eqnarray}
Then, the actual value of the ground state contribution at the BP stationary point becomes
\begin{eqnarray}
\label{BP-explicit-2} Z_{\mbox{BP;Fermi}}=Z_0(\gamma^{(\mbox{\small bp})})=\prod_{\{a,b\}\in{\cal G}_1}
\frac{H_{ab}H_{ba}\gamma_{ab}^{(\mbox{\small bp})}\gamma_{ba}^{(\mbox{\small bp})}}
{H_{ab}H_{ba}\gamma_{ab}^{(\mbox{\small bp})}\gamma_{ba}^{(\mbox{\small bp})}-1}\prod_{c\in{\cal
G}_{0}}\left(H_{cc}+\sum_{a'\sim c}(\gamma_{a'c}^{(\mbox{\small bp})})^{-1}\right),
\end{eqnarray}
where $\gamma^{(\mbox{\small bp})}$ is defined implicitly by Eqs.~(\ref{BP_eqs}).

\subsubsection{BP equations as no-loose-end constraints}
\label{subsub:ends}

We reiterate that the gauge fixing boils down to a particular choice for the set of parameters
$\{\gamma_{ab}\}_{a,b\in{\cal G}_{0};\{a,b\}\in{\cal G}_{1}}$. This can be done by imposing an
additional set of constraints that forbid the excited state structures with loose ends (more
precisely nodes of valence one). Note that loose ends that correspond to odd local excited states
are automatically forbidden due to the $\mathbb{Z}_{2}$-grading. Actually, this rationalizes our
choice of the skew-orthogonality constraints in
Eqs.~(\ref{skew-orthogonality},\ref{coeff-expressions}). Therefore, the BP conditions enforce a
cancellation of a large set of contributions to the partition function. The forbidden contributions
are those that contain loose end with even excited states at any vertex of the graph (while all
other edges adjusted to the vertex being in the ground state). Formalization of the BP constraints
results in
\begin{eqnarray}
\label{BP-equations} \forall a\in{\cal G}_0\ \ \mbox{and}\ \ \{a,c\}\in{\cal G}_1:\quad \left.\int
d{\bm\chi}_{a}d\bar{{\bm\chi}}_{a}f_{a}(\bar{{\bm\chi}}_{a},{\bm\chi}_{a})
e^{\bar{\chi}_{ca}\chi_{ca}/(H_{ac}H_{ca}\gamma_{ac})}\prod_{b\sim a}^{b\ne
c}e^{\gamma_{ba}\bar{\chi}_{ba}\chi_{ba}}\right|_{\gamma^{(\mbox{\small bp})}}=0.
\end{eqnarray}
The condition in Eq.~(\ref{BP-equations}) can also be restated as $\det(M'_{ac})=0$, where the
matrix $M'_{ac}$ is obtained from $M_{a}$ by replacing $\gamma_{ca}$ with
$\gamma'_{ca}=H_{ac}H_{ca}/\gamma_{ac}$. Utilizing Eq.~(\ref{det-M}) we observe that
Eqs.~(\ref{BP-equations}) turn explicitly into Eqs.~(\ref{BP_eqs}).

\subsubsection{Brief discussion of BP equations}

In the two preceding Subsections the BP equations were derived in two different ways, via the
variational and loose-end approaches respectively. Appendix \ref{sec:bose} also contain a relevant
information. It is shown there that the problem of finding the Fermi-BP-gauge (here we emphasize
that BP conditions follow from the Grassman/Fermi formulation) is completely equivalent to finding
a stationary point of the so-called Gaussian BP equations stated within the standard Gaussian
integrals. We call the standard BP approach Bose-BP to contrast it with the Fermi BP discussed
above.

Note, that BP equations can also be stated as defining extrema of the so-called Bethe Free Energy
functional, introduced for a general finite alphabet graphical model in \cite{05YFW}, and also
discussed in the context of the Gaussian (continuous alphabet) graphical model in \cite{08CH}.

\subsection{Loop Series for the Determinant}
\label{subsec:loop-series}

The loop series is obtained in a standard way by considering local excited states that correspond
to the choice of the second, third, or fourth term in Eq.~(\ref{skew-orthogonality}). Therefore,
each contribution to the correction for $Z_{0}$ determines a subgraph $C\subset{\cal G}$ that
consists of the edges on which excited states have been chosen. Due to the BP equations a subgraph
that provides a non-zero contribution does not have loose ends. Such subgraphs are referred to as
generalized loops. To describe a contribution fully we partition the edges of $C\subset{\cal G}$
into the neutral ones that correspond to the choice of the even local excited states and the
oriented edges that correspond to the odd local excited states. The choice of the third term in
Eq.~(\ref{skew-orthogonality}) will be denoted by an arrow on the edge $\{a,b\}$ that goes from $a$
to $b$, the choice of the fourth term is denoted by an opposite arrow that goes from $b$ to $a$.
The $\mathbb{Z}_{2}$ grading implies that in order to provide a non-zero contribution the number of
incoming arrows coincides with the number of the outgoing counterparts for all nodes of $C$.

We will further demonstrate that to provide a non-zero contribution each node of $C$ can actually
have no more than one incoming/outgoing arrow. This follows from the following form of the matrices
$M_{c}^{-1}$ inverse to $M_{c}$
\begin{eqnarray}
\label{M-inverse} (M_{c}^{-1})_{ab}=-\frac{1}{\gamma_{ac}\gamma_{bc}\left(H_{cc}+\sum_{a'\sim
c}(\gamma_{a'c})^{-1}\right)} \;\;\;\;\; {\rm for} \;\; a\ne b
\end{eqnarray}
and to their modified counterparts $M'_{c}$ obtained by replacing a certain number of
$\gamma_{ac}$-terms with $\gamma'_{ac}$-terms. We can further make use of the properties of the
Gaussian integrals and Eq.~(\ref{M-inverse}) to derive
\begin{eqnarray}
\label{corr-function} \int
d{\bm\chi}_{c}d\bar{\bm\chi}_{c}\bar{\chi}_{ac}\chi_{bc}e^{\sum_{a',a''\sim
c}M_{c,a'a''}\bar{\chi}_{a'c}\chi_{a''c}}=(M_{c}^{-1})_{ab}\det(M_{c})=-(H_{cc})^{-1}\prod_{a'\sim
c}^{a'\ne a,b}\gamma_{a'c} \;\;\; {\rm for} \;\; a\ne b.
\end{eqnarray}
The property described above follows from the fact
\begin{eqnarray}
\label{vanishing} \int
d{\bm\chi}_{c}d\bar{\bm\chi}_{c}\bar{\chi}_{a_{1}c}\ldots\bar{\chi}_{a_{k}c}\chi_{b_{1}c}\ldots\chi_{b_{k}c}
e^{\sum_{a',a''\sim c}M_{c,a'a''}\bar{\chi}_{a'c}\chi_{a''c}}=0 \;\;\; {\rm for} \;\;
\{a_{1},\ldots,a_{k}\}\cap \{b_{1},\ldots,b_{k}\}=\emptyset, \;\; k\ge 2.
\end{eqnarray}
To prove Eq.~(\ref{vanishing}) we apply the Wick's theorem that represents the Grassman integral in
Eq.~(\ref{vanishing}) as a sum of $k!$ contributions that correspond to $k!$ possible pairings
between the $\bar{\chi}$ and $\chi$ variables in the pre-exponents. Each contribution consists of a
product of $k$ pair correlation functions given explicitly by Eq.~(\ref{corr-function}). Due to the
form of the pair correlation functions, their product does not depend on a particular choice of the
pairing. On the other hand, the signs in front of the contributions are alternating and the number
of negative signs among the $k!$ contributions is the same as the number of positive signs,
provided $k\ge 2$. This leads to Eq.~(\ref{vanishing}). For $k=1$ the (non-zero) result is
given by Eq.~(\ref{corr-function}). These results apply as well to the modified
matrices $M'_{c}$.

\begin{figure*}
\centering
\vspace{0.1in}
\includegraphics[width=0.9\textwidth]{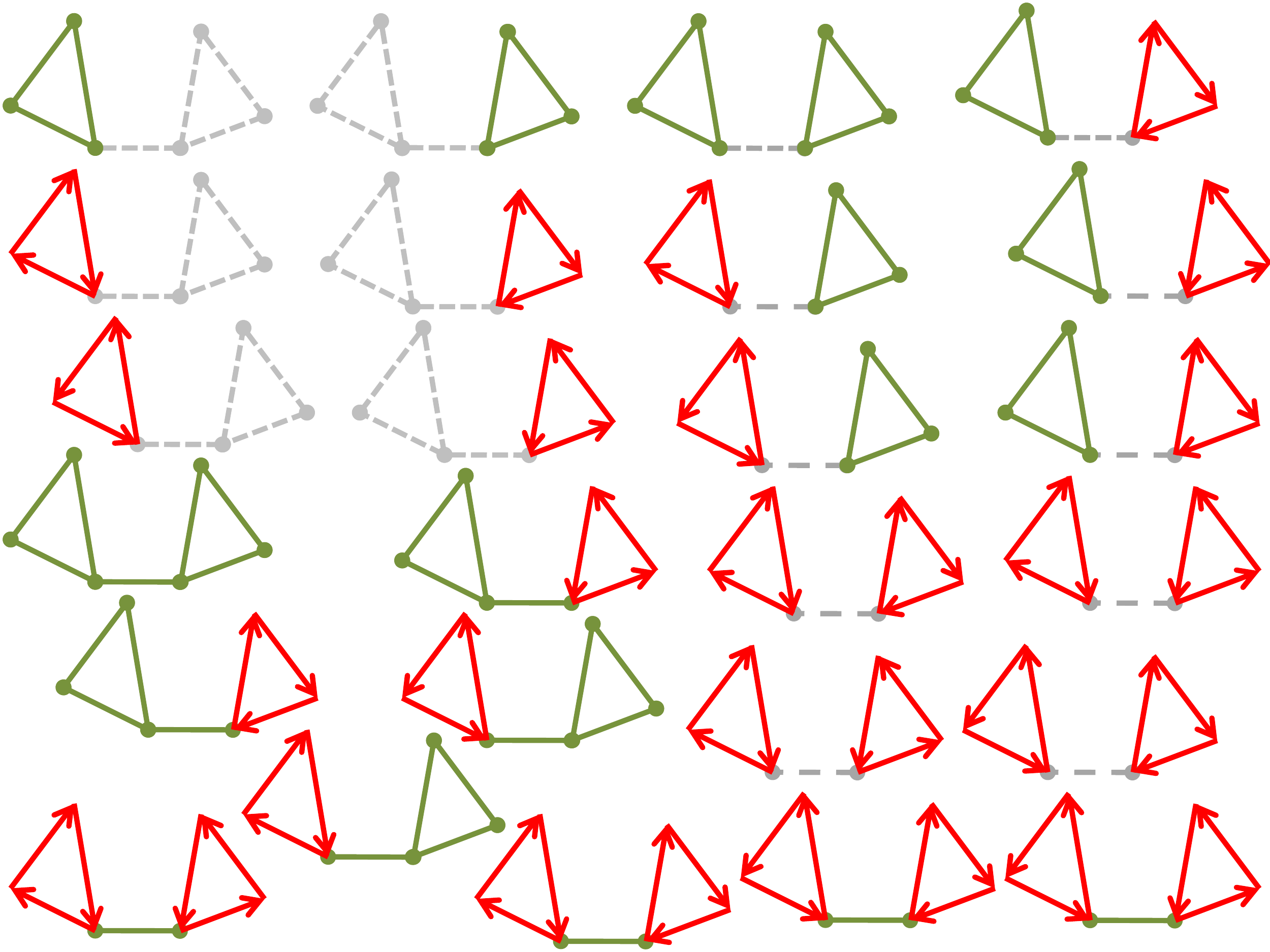}
\caption{Example of the set of generalized loops for a sample graph, consisting of $6$ vertexes and
$7$ edges. Generalized loops are shown in solid lines,  while part of the original graph which does
not belong to a generalized loop is shown in dashed gray. Oriented disjoint circles, which are also
subgraphs of respective generalized loops, are marked with arrows and shown in red. } \label{loops}
\end{figure*}

We are now in a position to summarize the results of
Eqs.~(\ref{M-inverse},\ref{corr-function},\ref{vanishing}) in the finite loop series expression
\begin{eqnarray}
\label{LS} \det(H)=Z=Z_{\mbox{BP;Fermi}}\left(1+\sum_{C\in GL({\cal G})}\sum_{C'\in
DOC(C)}r(C,C')\right)
\end{eqnarray}
where $GL({\cal G})$ is the set of generalized loops of the graph ${\cal G}$ (subgraphs with all
nodes of valence $2$ or higher), and $DOC(C)$ is the set of Disjoint Oriented Cycles (DOC) of the graph
$C$, i.e., subgraphs of $C$ whose all nodes have the valence exactly $2$, equipped with
orientation. For a DOC $C'$ let ${\rm deg}(C')={\rm card}(\pi_{0}(C'))$ be the
number of its connected components and for an edge $\alpha\in C'_{1}$ let $\partial_{0}\alpha \in
C'_{0}$ and $\partial_{1}\alpha \in C'_{0}$ be its left and right ends, respectively (orientation
arrows go from the left to the right). A relative contribution $r(C,C')$ can be represented as a
product of the node factors $r_{a}(C,C')$, edge factors $r_{\alpha}(C,C')$, and the parity factor:
\begin{eqnarray}
\label{r-as-product} r(C,C')=(-1)^{{\rm deg}(C')}\prod_{a\in C_{0}}r_{a}(C,C')\prod_{\alpha\in
C_{1}}r_{\alpha}(C,C')
\end{eqnarray}
where the factors are calculated in a straightforward way (see Appendix \ref{sec:node-edge-factors}
for the details)
\begin{eqnarray}
\label{r-factors} a\in C_{0}\setminus C'_{0}:\quad r_{a}(C,C')&=&\frac{H_{aa}+\sum_{a'\sim
a}^{a'\in C_{0}}H_{a'a}H_{aa'}\gamma_{aa'}^{(\mbox{\small bp})}+\sum_{a'\sim a}^{a'\in {\cal
G}_{0}\setminus C_{0}}(\gamma_{a'a}^{(\mbox{\small bp})})^{-1}}{H_{aa}+\sum_{a'\sim
a}(\gamma_{a'a}^{(\mbox{\small bp})})^{-1}}, \nonumber \\ \;\;\;\;\; a\in C'_{0}:\quad
r_{a}(C,C')&=&-\frac{1}{H_{aa}+\sum_{a'\sim
a}(\gamma_{a'a}^{(\mbox{\small bp})})^{-1}}, \nonumber \\
\alpha\in C_{1}\setminus C'_{1}:\quad r_{\alpha}(C,C')&=&-\frac{1}{H_{cd}H_{dc}
\gamma_{cd}^{(\mbox{\small bp})}\gamma_{dc}^{(\mbox{\small bp})}},
\nonumber\\
\alpha\in C'_{1}:\quad r_{\alpha}(C,C')&=&\frac{1-H_{cd}H_{dc}\gamma_{cd}^{(\mbox{\small bp})}
\gamma_{dc}^{(\mbox{\small bp})}}{H_{cd}\gamma_{cd}^{(\mbox{\small bp})}\gamma_{dc}^{(\mbox{\small bp})}},
\end{eqnarray}
where in the last two formulas $c=\partial_{0}\alpha$ and $d=\partial_{1}\alpha$.

\section{Summary and Conclusions}
\label{sec:summary}

The manuscript describes an explicitly constructed Loop Series for the Fermion Graphical Model. The
LS expresses determinant of an arbitrary square matrix in terms of a finite series. Four
important features of the series are:
\begin{itemize}

\item  The first term in the series corresponds to an approximation associated with solution of
the GBP equations, identical to those that emerged in the standard GGM discussed before
\cite{01WF,01RV,06MJW,03WJW,05RH,08MV,08CH,08SSWBD,08Joh}. Note, however, that while the standard
GGM approach would requires the matrix to be (at least) semi-definite (so that the normal variables
Gaussian integral would converge), our approach does not have this restriction as the Berezin
integrals are defined for any (even zero determinant!) square matrices.

\item Each term of the Loop Series is associated with a generalized loop and an oriented
disjoint cycle defined on the generalized loop.

\item Each term is expressed explicitly in terms of the chosen solution of the GBP equations.
Computation of any of the contributions (once the GBP solution is known) is the task of linear
complexity in the size of the underlying generalized loop.

\item The Loop Series can be constructed around any solution of the GBP equations, e.g., around
these solutions that are unstable with respect to the standard iterative GBP.
\end{itemize}

Let us now briefly discuss how the fermion based Loop Series could potentially be used in the
future. First of all, one hopes that the LS formula can clarify the accuracy of the GBP
approximation for different classes of matrices (sparse, walk-summable, planar, etc.). Second,
aiming to improve GBP one might be interested to identify problems (matrices) where accounting for
a relatively small, $O(N^\gamma)$ with $\gamma<3$, number of loops, will significantly improve the
GBP. In the context of these two general problems it will also be important to extend further
analysis of the Bethe Free energy for the Gaussian models initiated in \cite{05YFW,08CH}. The Bethe
Free energy is a functional whose stationary points coincide with the GBP solutions.

Finally, let us briefly overview the main idea of the second paper in the series \cite{08CCb}, and
its relation to the results discussed above. \cite{08CCb}  describes a construction generalizing LS
for determinant discussed in this paper. The construction starts with a $\mathbb{Z}_{2}$-gauge
theory, stated in terms of binary/Ising spins (that represent a gauge field with the discrete gauge
group $\mathbb{Z}_{2}$) and fermions on an arbitrary graph. It is shown that, on the one hand, the
graphical gauge model is reduced to a monomer-dimer model on the graph, while on the other hand it
turns into a series over disjoint oriented cycles on the graph, where the corresponding coefficient
is given by  determinant of a matrix related to to the full graph with excluded disjoint. We find
that this relation (between the monomer-dimer model and the Cycle Series) also follows (via a
certain type of inversion) from the Loop Series for the determinant discussed in this paper.

\section{Acknowledgments}

We are grateful to J. Johnson for useful comments. This material is based upon work supported by
the National Science Foundation under CHE-0808910. The work at LANL was carried out under the
auspices of the National Nuclear Security Administration of the U.S. Department of Energy at Los
Alamos National Laboratory under Contract No. DE-AC52-06NA25396.

\bibliographystyle{apsrev}
\bibliography{Gauss,BP_review}

\appendix

\section{Berezin Gaussian Integrals}
\label{sec:node-edge-factors}

In this technical Appendix we present some details of the calculations of the node $r_{a}(C,C')$
and edge $r_{\alpha}(C,C')$ factors for the relative loop contributions $r(C,C')$ given by
Eq.~(\ref{r-factors}). We start with defining the Grassman variables correlation function
\begin{eqnarray}
\label{corr-function-multi}
\left\langle\bar{\chi}_{a_{1}c}\ldots\bar{\chi}_{a_{k}c}\chi_{b_{1}c}\ldots\chi_{b_{k}c}
\right\rangle_{\bar{\chi},\chi}&=&Q_{c}^{-1}\int
d{\bm\chi}_{c}d\bar{\bm\chi}_{c}\bar{\chi}_{a_{1}c}\ldots\bar{\chi}_{a_{k}c}\chi_{b_{1}c}\ldots\chi_{b_{k}c}
e^{\sum_{a',a''\sim c}M_{c,a'a''}\bar{\chi}_{a'c}\chi_{a''c}} \nonumber \\ Q_{c}&=&\int
d{\bm\chi}_{c}d\bar{\bm\chi}_{c}e^{\sum_{a',a''\sim c}M_{c,a'a''}\bar{\chi}_{a'c}\chi_{a''c}}=\det(M_c).
\end{eqnarray}
The main feature of the Gaussian integral is expressed in the following Wick's theorem formula
\begin{eqnarray}
\label{Wick-theorem}
\left\langle\bar{\chi}_{a_{1}c}\ldots\bar{\chi}_{a_{k}c}\chi_{b_{1}c}\ldots\chi_{b_{k}c}\right\rangle
_{\bar{\chi},\chi}=\sum_{P}(-1)^{{\rm
deg}(P)}\prod_{j=1}^{k}\left\langle\bar{\chi}_{a_{j}c}\chi_{P(b_{j})c}\right\rangle_{\bar{\chi},\chi},
\end{eqnarray}
with the pair correlation function
$\left\langle\bar{\chi}_{ac}\chi_{bc}\right\rangle_{\bar{\chi},\chi}=(M_{c}^{-1})_{ab}$. Note, that
Eq.~(\ref{Wick-theorem}) follows naturally from a direct expansion of the following expression for
the generating function
\begin{eqnarray}
\label{generating-function} F(\bar{{\bm\xi}}_{c},{\bm\xi}_{c})=\left\langle e^{\sum_{a\sim
c}(\bar{\chi}_{ac}\xi_{ac}+\bar{\xi}_{ac}\chi_{ac})}\right\rangle_{\bar{\chi},\chi}
=e^{\sum_{a,b\sim c}(M_{c}^{-1})_{ab}\bar{\xi}_{ac}\xi_{bc}},
\end{eqnarray}
in powers of its argument.

The Loop Series is obtained upon the substitution of the skew-orthogonality condition
(\ref{skew-orthogonality}) into the integral representation (\ref{det=Z-vertex}) for the
determinant that results in the following expression
\begin{eqnarray}
\label{det-H-sum} \det(H)=Z_{\mbox{BP;Fermi}}+\sum_{C\in GL({\cal G})}\sum_{C'\in SL(C)}\prod_{\alpha\in{\cal
G}_{1}}H_{\partial_{0}\alpha,\partial_{1}\alpha}H_{\partial_{1}\alpha,\partial_{0}\alpha}
z_{\partial_{0}\alpha,\partial_{1}\alpha}(C,C')\prod_{a\in{\cal G}_{0}}(-H_{aa}z_{a}(C,C'))
\end{eqnarray}
The quantities $z_{ab}(C,C')$ with $a=\partial_{0}\alpha$, $b=\partial_{1}\alpha$  are determined
by the coefficients in the skew-orthogonal representation (\ref{skew-orthogonality}) and
are given by
\begin{eqnarray}
\label{z-alpha-expressions} z_{\alpha}(C,C')=c_{ab} \;\; {\rm for} \; \alpha\in {\cal
G}_{1}\setminus C_{1}, \;\;\; z_{\alpha}(C,C')=c_{ab}\kappa_{ab} \;\; {\rm for} \; \alpha\in
C_{1}\setminus C'_{1}, \;\;\; z_{\alpha}(C,C')=\zeta_{ab} \;\; {\rm for} \; \alpha\in C'_{1}
\end{eqnarray}
The explicit expressions are obtained by combining Eq.~(\ref{z-alpha-expressions}) with
Eq.~(\ref{coeff-expressions}). The quantities $z_{c}(C,C')$ are obtained by attaching the functions
of the local Grassman variables in (\ref{skew-orthogonality}) to the corresponding vertices
followed by the integration over the local Grassman variables. This results in
\begin{eqnarray}
\label{z-c-expressions}
c\in {\cal G}_{0}\setminus C_{0}:\quad
z_{c}(C,C')&=&\int d{\bm\chi}_{c}d\bar{{\bm\chi}}_{c}e^{\sum_{a,b\sim
c}M_{c,ab}\bar{\chi}_{ac}\chi_{bc}}=\det(M_{c})=\prod_{a\sim
c}\gamma_{a,c}(H_{cc})^{-1}\left(H_{cc}+\sum_{b\sim c}(\gamma_{bc})^{-1}\right), \nonumber \\
c\in C_{0}\setminus C'_{0}:\quad
z_{c}(C,C')&=&\int
d{\bm\chi}_{c}d\bar{{\bm\chi}}_{c}e^{\sum_{a,b\sim
c}M'_{c,ab}\bar{\chi}_{ac}\chi_{bc}}=\det(M'_{c}) \nonumber \\ &=&\prod_{a\sim c}^{a\in{\cal
G}_{0}\setminus C_{0}}\gamma_{a,c}\prod_{a\sim c}^{a\in
C_{0}}\gamma'_{a,c}(H_{cc})^{-1}\left(H_{cc}+\sum_{b\sim c}^{b\in{\cal G}_{0}\setminus
C_{0}}(\gamma_{bc})^{-1}+\sum_{b\sim c}^{b\in C_{0}}(\gamma'_{bc})^{-1}\right), \nonumber \\
c,s',s''\in C'_{0}:\quad z_{c}(C,C')&=&\int
d{\bm\chi}_{c}d\bar{{\bm\chi}}_{c}\bar{\chi}_{s''c}\chi_{s'c}e^{\sum_{a,b\sim
c}M'_{c,ab}\bar{\chi}_{ac}\chi_{bc}}=\det(M'_{c})(M'_{c})_{s''s'}^{-1} \nonumber
\\ &=&-(H_{cc})^{-1}\prod_{a\sim c}^{a\in{\cal G}_{0}\setminus C_{0}}\gamma_{ac}\prod_{a\sim c}^{a\in
C_{0}\setminus C'_{0}}\gamma'_{ac}.
\end{eqnarray}
The loop series (\ref{LS}) is obtained from the representation of Eq.~(\ref{det-H-sum}) with
the ingredients given by Eqs.~(\ref{z-alpha-expressions}) and (\ref{z-c-expressions}) by regrouping
the factors into the node and edge ones in a very obvious way.

Finally, we demonstrate that the vertices that have more than one incoming/outgoing arrow do not
contribute to the determinant. Such contribution is proportional to
$\left\langle\bar{\chi}_{a_{1}c}\ldots\bar{\chi}_{a_{k}c}\chi_{b_{1}c}\ldots\chi_{b_{k}c}\right\rangle
_{\bar{\chi},\chi}$ with $\{a_{1},\ldots,a_{k}\}\cap \{b_{1},\ldots,b_{k}\}=\emptyset$ and
vanishes, since applying the Wick's theorem Eq.~(\ref{Wick-theorem}) we derive
\begin{eqnarray}
\label{proof-vanishing}
\left\langle\bar{\chi}_{a_{1}c}\ldots\bar{\chi}_{a_{k}c}\chi_{b_{1}c}\ldots\chi_{b_{k}c}\right\rangle
_{\bar{\chi},\chi}&=&\sum_{P}(-1)^{{\rm
deg}(P)}\prod_{j=1}^{k}\left\langle\bar{\chi}_{a_{j}c}\chi_{P(b_{j})c}\right\rangle_{\bar{\chi},\chi}
=\sum_{P}(-1)^{{\rm deg}(P)}\prod_{j=1}^{k}(M'_{c})_{a_{j},P(b_{j})}^{-1} \nonumber \\ &=&
\frac{-\sum_{P}(-1)^{{\rm deg}(P)}}{(H_{cc}+\sum_{a'\sim c}^{a'\in{\cal G}_{0}\setminus
C_{0}}(\gamma_{a'c})^{-1}+\sum_{a'\sim c}^{a'\in
C_{0}})(\gamma'_{a'c})^{-1})\prod_{j=1}^{k}(\gamma_{a_{j}c}\gamma_{b_{j}c})}=0,
\end{eqnarray}
and the alternating sum over $k!$ permutations $P$ in the numerator vanishes for $k\ge 2$.

\section{Belief-Propagation Gauge in Bose-representation}
\label{sec:bose}

In the normal (Bose) integral representation the determinant can be represented as follows
\begin{eqnarray}
(\det H)^{-1}=\int \prod_a
\left(\frac{d\bar{\psi}_ad\psi_a}{2\pi}\right)\exp\left(-\frac{1}{2}\sum_{a\in{\cal
G}_0}H_{aa}\bar{\psi}_a\psi_a-\frac{1}{2}\sum_{(a,b)\in{\cal G}_1}H_{ab}\bar{\psi}_a\psi_b\right),
\label{det-bose}
\end{eqnarray}
where the integration goes over the pairs of the complex conjugated  variables (normal
$c$-numbers). Here, we restrict ourselves to the case when the integral is well-defined
(convergent). To decouple the terms associated with different vertices of ${\cal G}_0$ we introduce
the following Bose-version of the HS transformation
\begin{eqnarray}
 && \exp\left(-\frac{H_{ab}}{2}\bar{\psi}_a\psi_b-\frac{H_{ba}}{2}\bar{\psi}_b\psi_a\right)=
 -\frac{1}{(2\pi)^2H_{ab}H_{ba}}\int d\bar{\varphi}_{ab}d\varphi_{ab}d\bar{\varphi}_{ba}d\varphi_{ba}
 \exp\left(-\frac{H_{ab}^{-1}}{2}\bar{\varphi}_{ab}\varphi_{ba}-
 \frac{H_{ba}^{-1}}{2}\bar{\varphi}_{ba}\varphi_{ab}\right)\nonumber\\
 && \times \exp\left(i\varphi_{ba}\bar{\psi}_a/2+i\bar{\varphi}_{ba}\psi_a/2+
 i\varphi_{ab}\bar{\psi}_b/2+i\bar{\varphi}_{ab}\psi_b/2\right).
 \label{HSb}
\end{eqnarray}
Substituting Eq.~(\ref{HSb}) into Eq.~(\ref{det-bose}) and performing the integration over the
vertex variables $\psi$ we arrive at the following Gaussian edge factor function formulation
\begin{eqnarray}
 && (\det H)^{-1}=(2\pi)^{-2|{\cal G}_1|}\left(\prod_{a\in{\cal G}_0} H_{aa}^{-1}\right)
 \left(\prod_{(a,b)\in{\cal G}_1} (-H_{ab}H_{ba})^{-1}\right)
 \int{\cal D}{\bar{\bm\varphi}}{\cal D}{\bm \varphi}\prod_{a\in{\cal G}_0}
 F_a\left(\bar{\bm \varphi}_a,{\bm \varphi}_a\right)\prod_{\alpha\in{\cal G}_1}
 G_\alpha\left(\bar{\bm \varphi}_\alpha,{\bm \varphi}_\alpha\right),
\label{GB1}\\
 && F_a\left(\bar{\bm \varphi}_a,{\bm \varphi}_a\right)\equiv
 \exp\left(-\frac{H_{aa}^{-1}}{2}\left(\sum_{b\sim a}\varphi_{ba}\right)
 \left(\sum_{b'\sim a}\bar{\varphi}_{b'a}\right)\right),
 \label{GB2}\\
 && G_\alpha\left(\bar{\bm \varphi}_\alpha,{\bm \varphi}_\alpha\right)
 =\exp\left(-\frac{H_{ab}^{-1}}{2}\bar{\varphi}_{ab}\varphi_{ba}-
 \frac{H_{ba}^{-1}}{2}\bar{\varphi}_{ba}\varphi_{ab}\right).
 \label{GB3}
\end{eqnarray}
We further introduce the following gauge representation for the skew scalar product
\begin{eqnarray}
 G_\alpha=\phi_{ab}\exp\left(-\frac{\lambda_{ab}}{2}\bar{\varphi}_{ab}\varphi_{ab}-
 \frac{\lambda_{ba}}{2}\bar{\varphi}_{ba}\varphi_{ba}\right)+
 \sum_{n=1}^\infty Q_{ab}^{(n)}(\bar{\varphi}_{ab},\varphi_{ab})
 Q_{ba}^{(n)}(\bar{\varphi}_{ba},\varphi_{ba}),
 \label{G-skew-B}
\end{eqnarray}
where $\phi_{ab}=\phi_{ba}$; the first term on the rhs of Eq.~(\ref{G-skew-B}) corresponds to the
local ground state while the sum (remainder) represents some discrete spectrum of the excited
states. The remaining freedom in Eq.~(\ref{G-skew-B}) is fixed via the following BP-(zero loose
end) conditions
\begin{eqnarray}
 \forall \{a,b\}\in{\cal G}_1:\quad
 \int{\cal D}\bar{\bm\varphi}_a {\cal D}{\bm\varphi}_a F_a(\bar{\bm \varphi}_a,{\bm\varphi}_a)
 \left(\prod_{c\sim a}^{c\neq b}\exp\left(-\frac{\lambda_{ca}}{2}\bar{\varphi}_{ca}\varphi_{ca}\right)\right)
 Q_{ba}^{(n)}(\bar{\varphi}_{ba},\varphi_{ba})=0.
\label{BP-gauge}
\end{eqnarray}
Multiplying Eqs.~(\ref{BP-gauge}) with $Q_{ba}^{(n)}(\bar{\varphi}_{ba},\varphi_{ba})$, making
summation over all excited states ($n>0$), and expressing the $Q$-terms via the ground state
contribution (with the help of Eqs.~(\ref{G-skew-B})), one arrives at the following BP relations
stated solely in terms of the local ground states (local gauges)
\begin{eqnarray}
 && \forall a,b\in{\cal G}_0,\ a\sim b:\quad
\int{\cal D}\bar{\bm\varphi}_a {\cal D}{\bm\varphi}_a F_a(\bar{\bm \varphi}_a,{\bm\varphi}_a)
 \left(\prod_{c\sim a}^{c\neq b}\exp\left(-\frac{\lambda_{ca}}{2}\bar{\varphi}_{ca}\varphi_{ca}\right)\right)
 \exp\left(-\frac{H_{ab}^{-1}}{2}\bar{\varphi}_{ab}\varphi_{ba}-
 \frac{H_{ba}^{-1}}{2}\bar{\varphi}_{ba}\varphi_{ab}\right)\nonumber\\ && =\phi_{ab}
 \exp\left(-\frac{\lambda_{ab}}{2}\bar{\varphi}_{ab}\varphi_{ab}\right)
 \int{\cal D}\bar{\bm\varphi}_a {\cal D}{\bm\varphi}_a F_a(\bar{\bm \varphi}_a,{\bm\varphi}_a)
 \left(\prod_{c'\sim a}\exp\left(-\frac{\lambda_{c'a}}{2}\bar{\varphi}_{c'a}\varphi_{ac'}\right)\right).
\label{rel-gauge}
\end{eqnarray}
Evaluation of the Gaussian integrals transforms Eqs.~(\ref{rel-gauge}) into
\begin{eqnarray}
 \forall a,b\in{\cal G}_0,\ a\sim b:\quad  -\lambda_{ab}H_{ab}H_{ba}&=&H_{aa}
 +\sum_{c\sim a}^{c\neq b}(\lambda_{ca})^{-1},\label{rel-gaugeA}\\
 \phi_{ab}&=&\lambda_{ba}\left(H_{aa}+\sum_{c\sim a}(\lambda_{ca})^{-1}\right)
 =1-\lambda_{ab}\lambda_{ba}H_{ba}H_{ab},
\label{rel-gaugeB}
\end{eqnarray}
where the transformation from the lhs of Eq.~(\ref{rel-gaugeB}) to its rhs  also involves
Eqs.~(\ref{rel-gaugeA}). The resulting BP expression for the entire partition function (BP
expression for the inverse determinant) becomes
\begin{eqnarray}
 && Z_{\mbox{BP;Bose}}=(2\pi)^{-2|{\cal G}_1|}\left(\prod_{a\in{\cal G}_0} H_{aa}^{-1}
 \int{\cal D}\bar{\bm\varphi}_a{\cal D}{\bm \varphi}_a
 F_a\left(\bar{\bm \varphi}_a,{\bm \varphi}_a\right)\exp\left(-\sum\limits_{b\sim a}
 \frac{\lambda_{ba}}{2}\bar{\varphi}_{ba}\varphi_{ba}\right)\right)
 \left(\prod_{\{a,b\}\in{\cal G}_1}\phi_{ab}/(-H_{ab}H_{ba})\right)\nonumber\\
 && =\frac{\prod_{\{a,b\}\in{\cal G}_1}\left(1-(\lambda_{ab}\lambda_{ba}H_{ab}H_{ba})^{-1}\right)
 }{\prod_{a\in{\cal G}_0}\left(H_{aa}+\sum_{c'\sim a}(\lambda_{c'a})^{-1}\right)}.\label{Z_BP_Bose}
\end{eqnarray}
Comparing Eq.~(\ref{Z_BP_Bose}) with Eq.~(\ref{BP-explicit-2}) we conclude that BP expressions in
the Fermi- and Bose- cases are fully equivalent, i.e.
\begin{eqnarray}
Z_{\mbox{BP;Bose}}\cdot Z_{\mbox{BP;Fermi}}=1.\label{Super}
\end{eqnarray}
To summarize, we have shown in this Appendix that the Bose-BP equations of the standard Gaussian
Graphical model, e.g., discussed in \cite{01WF,01RV,06MJW,03WJW,05RH,08MV,08CH,08SSWBD,08Joh}, are
equivalent to the ones derived within the Fermi-BP approach discussed in the main part of the text.
Moreover, the BP partition functions of the Fermi and Bose models are inversely proportional to
each other.

One final remark of this Appendix (and of the manuscript) concerns the reconstruction of the full
expression for the inverse determinant from BP. The ``ground state" part of the BP-gauge is fully
defined by Eqs.~(\ref{rel-gaugeA},\ref{rel-gaugeB}), however, a freedom in calculating the
``excited state" gauges $Q^{(n)}$ still remains. The $Q^{(n)}$ gauges can be fixed in the spirit of
\cite{07CC}, resulting in an infinite loop tower hierarchy for the inverse determinant. Note, that
this infiniteness is in contrast with the finiteness of the Loop Series, built in the main part of
the manuscript for the determinant of the same matrix. Here we do not discuss the Loop Tower
reconstruction for the inverse determinant, leaving the question of possible relation between the
aforementioned infinite and finite series open for future investigations.

\end{document}